\documentclass[aps,prl,reprint,superscriptaddress, showpacs]{revtex4-1}

\usepackage{graphicx}
\usepackage{dcolumn}
\usepackage{bm}
\usepackage{xcolor}
\usepackage{longtable}

\begin{document}


\title{Experimental demonstration of shaken lattice interferometry}


\author{C.A. Weidner}
\affiliation{Department of Physics and JILA, University of Colorado, Boulder, Colorado, 80309-0440, USA}
\author{Dana Z. Anderson}
\affiliation{Department of Physics and JILA, University of Colorado, Boulder, Colorado, 80309-0440, USA}
\email{dana@jila.colorado.edu}


\date{\today}

\begin{abstract}
{We experimentally demonstrate a shaken lattice interferometer. Atoms are trapped in the ground Bloch state of a red-detuned optical lattice. Using a closed-loop optimization protocol based on the dCRAB algorithm, we phase-modulate (shake) the lattice to transform the atom momentum state. In this way, we implement an atom beamsplitter and build five interferometers of varying interrogation times $T_\mathrm{I}$. The sensitivity of shaken lattice interferometry is shown to scale as $T_\mathrm{I}^2$, consistent with simulation \cite{Anderson2017}. Finally, we show that we can measure the sign of an applied signal and optimize the interferometer in the presence of a bias signal.}
\end{abstract}


\maketitle


The wavefunction describing an ensemble of atoms in an optical lattice will evolve when the lattice is subjected to amplitude and/or phase modulation. The work of Ref. \cite{Meystre2001} showed that a prescribed final state wavefunction can be obtained from an intial known wavefunction by genetic optimization of a time-dependent phase modulation; i.e., by learning how to appropriately shake the lattice. In Ref. \cite{Anderson2017} we extended this pioneering idea, showing numerically that one can utilize a shaken lattice to perform atom interferometry. Utilizing a series of shaking protocols the quantized momentum states of the atoms in an optical lattice are transformed and made to undergo a conventional interferometry sequence of splitting, propagation, reflection, reverse propagation, and recombination. Configured as a Michelson interferometer \cite{Wu2005} the shaken lattice interferometer was shown to be sensitive to inertial forces with the same $T_\mathrm{I}^2$ dependence on interrogation time \cite{Anderson2017} as free-space atom interferometers \cite{Kasevich2013, Robins2014}. More generally the shaken lattice approach allows tuning of the interferometer transfer function, e.g. to minimize sensitivity to a constant acceleration.

This work presents the first experimental demonstration of a shaken-lattice interferometer. Optical lattices have been used to accelerate atoms in interferometers via Bloch oscillations \cite{Robins2013, Tino2016}. Shaken optical lattices have been used to measure gravity \cite{Tino2011, Tino2012}. Our method differs in that we shake the lattice to transform the atom wavefunction from an initial state to a desired final state. While this method may be more generally applied to control an atom wavefunction, we build an atom-based inertial sensor with square-law dependence on $T_\mathrm{I}$. We demonstrate that the shaken lattice interferometer is capable of sensing acceleration signals; notably, the sign of the acceleration signal is readily distinguished. As a first step towards the demonstration of the tunable sensitivity predicted in Ref. \cite{Anderson2017} we optimize the interferometer in the presence of a bias signal and show sensitivity to perturbations on this bias.

To find the shaking protocol that best performs the necessary state-to-state transformations, we perform gradient-free, closed-loop optimization based on the dCRAB algorithm \cite{Montangero2011, Montangero2015}. Other experiments have used this algorithm to optimize the state inversion of a BEC \cite{Hohenester2013, Montangero2016}, in Ramsey interferometry schemes \cite{Schmiedmayer2014}, or to calibrate qubit operations in diamond NV centers \cite{Jelezko2017}. Optimization protocols have also been used in cold atom \cite{Schumm2008, Hush2015} and quantum optics experiments \cite{Zeilinger2016} as well as to find efficient pulse schemes in light-pulse atom interferometry \cite{Close2012}.

\begin{figure}[hb]
\includegraphics[scale = .52]{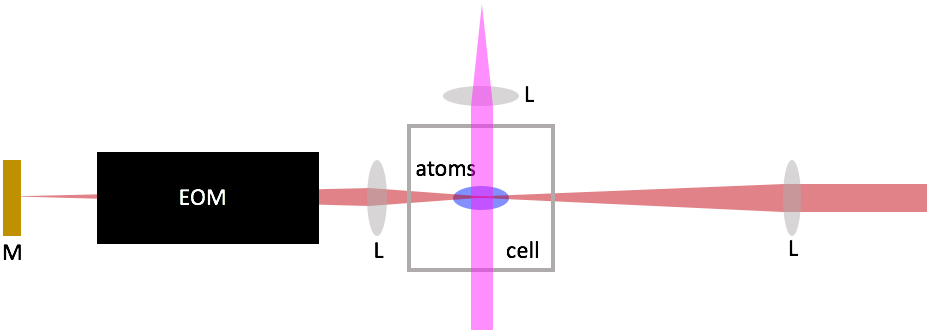}
\caption{\label{fig:exptscheme} Top view of the experimental setup. The optical lattice beam (dark red) propagates through a lens (L) and focuses on the atoms in the center of the vacuum cell. A cat's eye system placed after the atoms passes the beam through an EOM and focuses the beam onto a retro-reflecting mirror. The imaging beam (magenta) images the atoms' momentum state after being dropped from the lattice and falling for $20$~ms time-of-flight.}
\end{figure}

Our experiment is based on the compact BEC setup described in Ref. \cite{Anderson2010} and shown schematically in Fig. \ref{fig:exptscheme}. We trap $^{87}$Rb atoms in the $|F, m_\mathrm{F}\rangle = |2,2\rangle$ state on an atom chip and cool them to degeneracy via forced RF evaporation. Similar atom chip-based systems have been used to build compact optical lattice systems \cite{Salim2015} and study atomtronics \cite{Anderson2016}. The condensed atoms are loaded into the ground Bloch state \cite{Phillips2002} of a red-detuned optical lattice with a depth $V_0 \approx 14E_\mathrm{r}$, where the recoil energy is $E_\mathrm{r} = \hbar^2 k_\mathrm{L}^2/2m$ for an atom mass $m$ and a lattice wavenumber $k_\mathrm{L} = 2\pi/\lambda_\mathrm{L}$.

The optical lattice is formed by retro-reflecting a laser of wavelength $\lambda_\mathrm{L} = 852$~nm. The lattice light is locked to the $|F = 4\rangle\rightarrow |F' = 4/5\rangle$ crossover transition in cesium and comes to a focus ($w_0 \approx 40$~$\mathrm{\mu}$m) at the atoms' position in the vacuum cell. The beam passes twice through an electro-optic modulator (EOM) placed after the cell. The EOM shifts the phase of the reflected light relative to the incident light; the lattice is shaken by modulating the voltage applied to the EOM crystal. The desired shaking function is generated using an arbitrary waveform generator (AWG, Agilent 33250A), amplified by a factor of $40$, and fed into the EOM input. We calibrate the phase change as a function of the AWG output using an optical Michelson interferometer and obtain $0.746(6)$~rad/V, where the parentheses give the error in the last digit. Shaking the lattice diffracts atoms into quantized momentum states separated by $2\hbar k_\mathrm{L}$ \cite{Meystre2001, Anderson2017}. We image the atoms' momentum state populations via standard time-of-flight absorption imaging. This allows for optimization of the interferometer sequence and calibration of the atoms' response to an applied signal.

To calibrate the atoms' response to an applied signal, a pair of coils outside the cell provides a magnetic field gradient $G = \partial B/\partial x$ along the lattice direction \footnote{A gradient is also applied in the direction orthogonal to the lattice direction (and to gravity), but the only effect is a negligible shift in the lattice trap position along this direction.}. A bias magnetic field remains on while the atoms are trapped in the lattice to maintain the atoms' spin polarization and thus their magnetic field sensitivity. The gradient $G$ gives rise to an effective acceleration
\begin{equation}
\label{eq:aeff}
a_\mathrm{eff} = G g_\mathrm{F}m_\mathrm{F}\mu_\mathrm{B}/m
\end{equation}
where $g_\mathrm{F} = 1/2$ is the Land{\'e} g-factor \cite{SteckRb87}, and $\mu_\mathrm{B}$ is the Bohr magneton. In practice we calibrate the acceleration due to the gradient field by loading the atoms into a dipole trap and measuring their velocity as a function of hold time in the trap while varying $G$. The applied acceleration $a_\mathrm{eff}$ increases linearly with current through the gradient coils ($a_\mathrm{eff} = 0.71\pm 0.16$~m/s$^2$/A), as expected from calculations using the Biot-Savart law.

To build the interferometer sequence the shaking function is optimized to provide the desired state-to-state transformations. In particular we ``stitch'' together shaking functions corresponding to different interferometer operations \cite{Anderson2017}. For example, to implement an atom beamsplitter we load atoms into the ground Bloch state of the lattice. The lattice is subsequently shaken to split the atom wavefunction so that roughly half of the atoms occupy each of the $\pm 2\hbar k_\mathrm{L}$ momentum states. We then optimize separate propagation protocols that maintain the split state. To recombine the atoms back into the ground state (in the absence of an applied signal) the optimized splitting shaking protocol is run in reverse. Each protocol is $T = 0.2$~ms in duration and is multiplied by an envelope function $f_\mathrm{env}(x) = \sin^2(\pi t/T)$ to ensure smooth turn-on and turn-off of the shaking. This allows the shaking functions to be stitched together without discontinuity. In this way we optimize five separate interferometers with interrogation times of $T_n = 0.4n$~ms, where $n = 1,..,5$. The splitting and recombination times are included in the definition of the total interrogation time because they are not negligibly small relative to the propagation time.

\begin{figure}[ht]
\includegraphics[scale = .335]{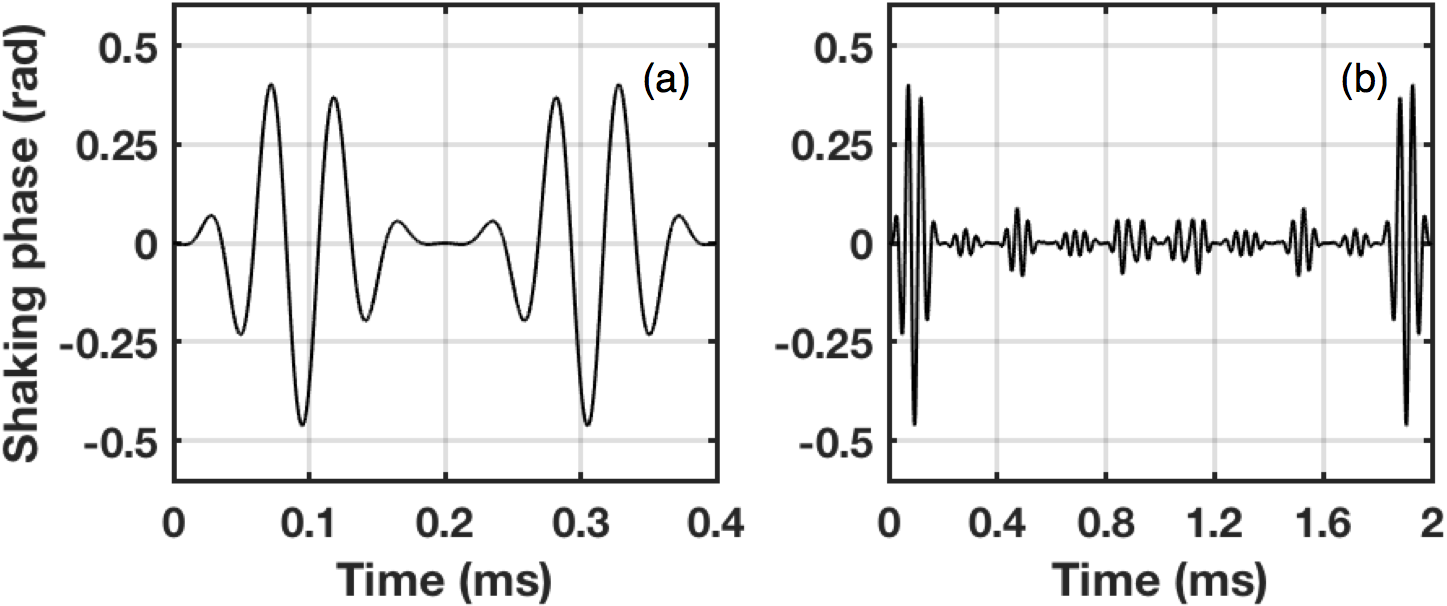}
\caption{\label{fig:shaking} Two example shaking functions. (a) Splitting and recombination shaking protocols. (b) Protocol from (a), but with 4 propagation steps added after splitting. We optimize the interferometer so that the atoms remain split at the end of each propagation step. In both cases the second half of the shaking protocol is simply the reflection of the first half.}
\end{figure}

To optimize the interferometer we define a the split state as our target state. The dCRAB algorithm picks five frequencies at random within our chosen frequency band of $18-30$~kHz \footnote{This frequency range is chosen because it surrounds the Bloch band $0\rightarrow 1$ transition. The atoms' momentum population changes most dramatically in this band. This narrows our search space for faster convergence on the desired state.} and assigns each frequency five separate Fourier sine and cosine amplitudes. The five waveforms described by these Fourier coefficients become the five vertices of a simplex in frequency space. Using the Nelder-Mead algorithm, the simplex is modified and iteratively converges upon the target state. Error is determined by building a vector $\vec{P}$ with components $P_n$ containing the relative population of atoms in the $2n\hbar k_\mathrm{L}$ momentum states. In practice there is negligible population in the $\pm 6\hbar k_\mathrm{L}$ states, so $|n|$ is truncated to $N = 2$. The percent error $E$ is then defined as
\begin{equation}
\label{eq:err}
E = \bigg (1-\frac{\vec{P}\cdot\vec{P}_\mathrm{des}}{\big |\vec{P}\big | \big |\vec{P}_\mathrm{des} \big |} \bigg )\times 100\%
\end{equation}
where $\vec{P}_\mathrm{des}$ is the vector corresponding to the desired state. For example, the desired momentum state vector for splitting is $P_\mathrm{des,~sp} = (0,~0.5,~0,~0.5,~0)$. Two examples of optimized shaking functions are shown in Fig. \ref{fig:shaking}. While splitting requires relatively high shaking amplitudes, smaller amplitudes are required to maintain this state during propagation. This is likely because the split state is similar to the first excited Bloch state of the lattice, so less modulation is required to maintain a state close to a lattice eigenstate than to transform from one state to another nearly orthogonal state.

\begin{figure}[ht!]
\includegraphics[scale = .74]{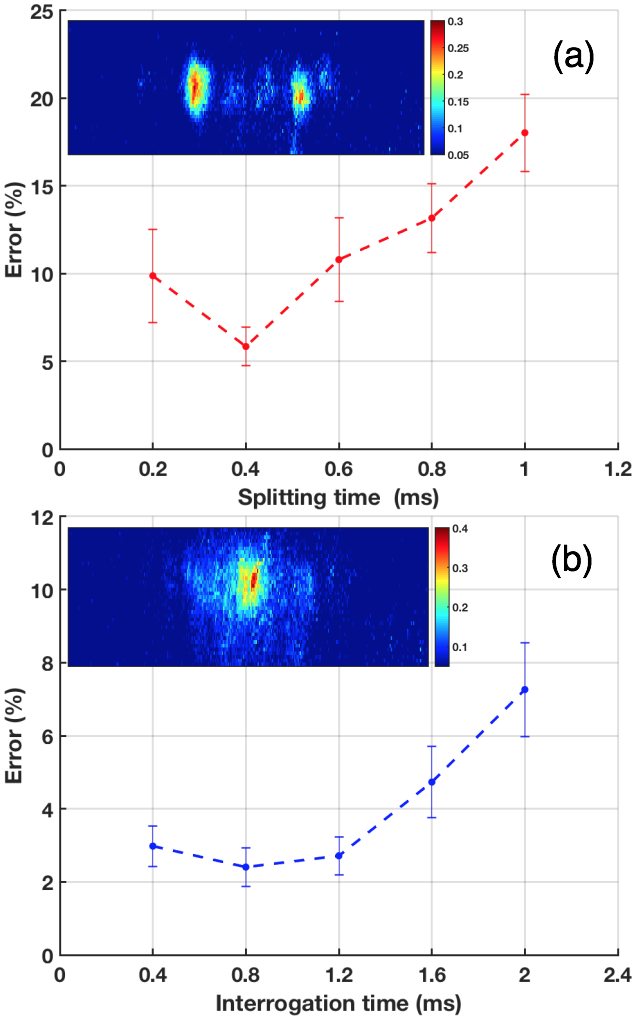}
\caption{\label{fig:sprec} Percent error in the (a) splitting and (b) recombination protocols as a function of (a) the splitting time $T_\mathrm{I}/2$ and (b) the total interrogation time $T_\mathrm{I}$. (a, inset) An image from optimized splitting of the atoms into equal population in the $\pm 2\hbar k_\mathrm{L}$ momentum states and (b, inset) recombining the atoms into the ground Bloch state. The colorbar represents optical density (OD).}
\end{figure}

The error in splitting begins at roughly $10\%$ and increases as propagation protocols are added (Fig. \ref{fig:sprec}a). This error arises due to spurious atoms detected in the $0\hbar k_\mathrm{L}$ momentum state due to atom localization in the deep lattice potential \cite{Raithel2009} and atom-atom interactions causing heating and loss of visibility during the experiment \cite{Raithel2009, Bloch2017, MuellerE2015, MuellerE2015b} or during time-of-flight \cite{Esslinger2001}. We are also limited by asymmetry between the two split clouds and the finite momentum spread of the condensed atoms as they are loaded into the lattice. Simulations show that the momentum spread of the atoms limits the error to about $1\%$. Upon recombination our errors are $<10\%$ (Fig. \ref{fig:sprec}b). Errors in recombination manifest largely as population of higher-order momentum states due to accumulated errors in the splitting and propagation protocols. The error in recombination is lower than splitting because spurious atoms detected in the $0\hbar k_\mathrm{L}$ momentum state are no longer deleterious.

To quantify the performance of the interferometer we measure how the final momentum state vector $\vec{P}_a$ changes as an acceleration signal $a$ is applied to the atoms. As determined in Ref. \cite{Anderson2017}, because there are more than two momentum states considered here, we cannot assign a phase difference based on the relative path length traveled by two arms of the interferometer. Thus, we use the classical Fisher information (CFI) to define the lowest detectable acceleration $\delta a$ based on the Cramer-Rao (CR) bound \cite{Haine2016}. We can define the CFI as \cite{Anderson2017}
\begin{equation}
\label{eq.FI_class_P}
F_{C,P}(a) = N_\mathrm{at}\sum_{n=-N}^N\,\frac{(\partial{P_{\mathrm{a},n}/\partial a)^2}}{P_{\mathrm{a},n}} = N_\mathrm{at}(\vec{A}\cdot\vec{B})
\end{equation}
where $\vec{A}$ has components $A_n = 1/P_{\mathrm{a}, n}$ and $\vec{B}$ has components $B_n = (\partial P_{\mathrm{a}, n}/\partial a)^2$ and $N_\mathrm{at}$ is the total atom number. We numerically evaluate derivatives using a two-point forward-difference scheme. The CR bound allows us to find the minimum detectable acceleration $\delta a = 1/\sqrt{F_{C,P}}$.

\begin{figure}[ht]
\includegraphics[scale = .215]{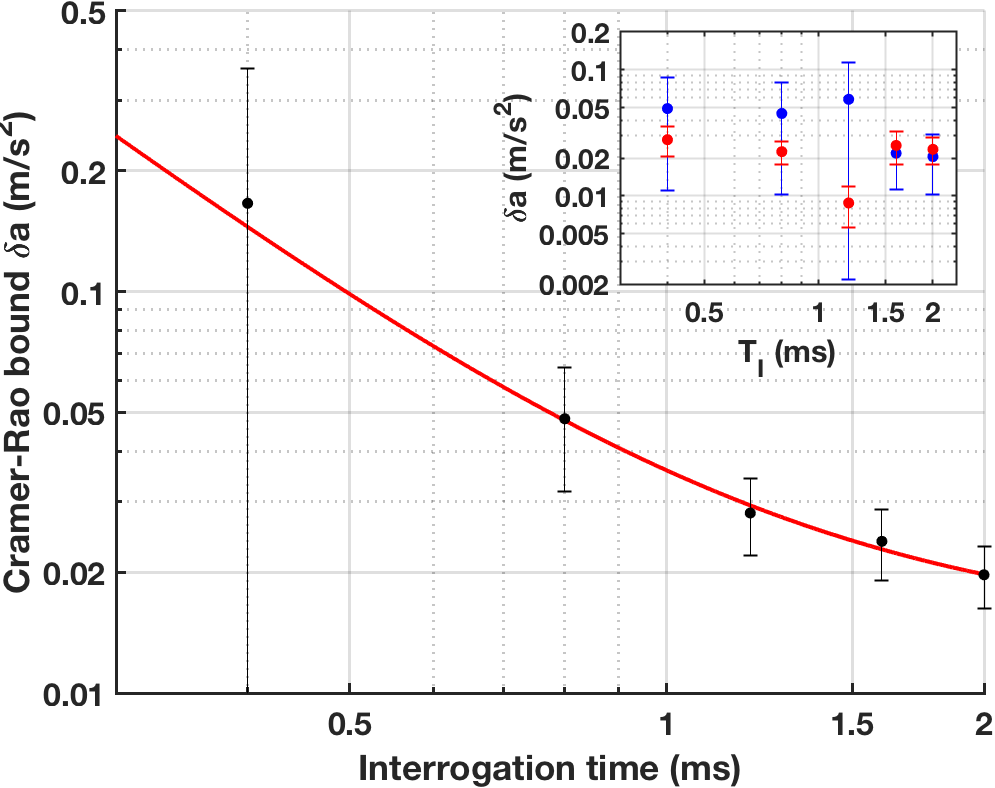}
\caption{\label{fig:scaling} Minimum detectable effective acceleration $\delta a$ plotted as a function of interrogation time for each of the five interferometers optimized for this work (black) and fit (red) to $f(T_\mathrm{I}) = aT_\mathrm{I}^b + c$. The scaling value is $b$ is consistent with the expected $T_\mathrm{I}^2$ scaling, and the offset $c$ arises due to imaging noise and is measured experimentally. (inset) Data taken with no atoms present (blue) and no shaking applied to the atoms (red) showing no signal other than imaging noise. Blue data is scaled by the ratio of the relative atom numbers as explained in the text.}
\end{figure}

Results of performing this analysis are shown in Fig. \ref{fig:scaling}. The data is fit using the Levenberg-Marquardt scheme \cite{Levenberg1944, Marquardt1963} to a function $f(T_\mathrm{I}) = aT_\mathrm{I}^{-b} + c$ where $b$ is the sensitivity scaling and $c$ is a noise-limited offset that we can quantify. Therefore we fit only the values of $a$ and $b$. To mitigate the effects of imaging noise we set a threshold OD below which we do not count atoms. We find that the optimum value of this threshold is $OD_\mathrm{thresh} \approx 0.05-0.06$ depending on the imaging noise. The largest contributor to this noise is imbalance in the exposure time between the absorption and background images.

This offset $c$ is measured by ``measuring'' the CR bound without atoms present, then dividing this number by the ratio of the detected atom number with and without atoms actually being present. Future work will focus on the reduction of this offset by improving the exposure balance in the imaging system. The atom signal-to-noise ratio can be improved by minimizing the heating of the atoms in the lattice \cite{Greiner2015}, which will also allow for longer interrogation times. However, longer interrogation times will increase decoherence due to phase diffusion \cite{Java1997}, which can be mitigated by lowering the atom density in the lattice.

Our fit (Fig. \ref{fig:scaling}) gives $b = 1.96 \pm 0.13$ using our measured value of $c = 0.014(3)$, consistent with the expected $T_\mathrm{I}^2$ scaling. The measured value for $c$ and the fit for $b$ are consistent with our results when we leave both $b$ and $c$ to be free parameters; in this case, we measure $b = 2.20\pm 0.34$ and $c = 0.015(2)$. Data taken on different days gives scaling that is consistent with the expected $T_\mathrm{I}^2$ scaling, and the data presented here is a typical example. Furthermore, data taken where a signal is applied to unshaken atoms is indistinguishable from noise and shows no discernable scaling law, showing that the shaking is a coherent process \footnote{For the accelerations presented here, the interrogation time is much less than the Bloch oscillation time $\tau_\mathrm{B} \propto 1/a$. Thus we do not expect the atoms population to change dramatically when accelerated without shaking. Due to the $T_\mathrm{I}^2$ scaling, as we increase $T_\mathrm{I}$, the lower values of $a$ that we are sensitive to strengthen this assumption.}.

\begin{figure}[ht]
\includegraphics[scale = .23]{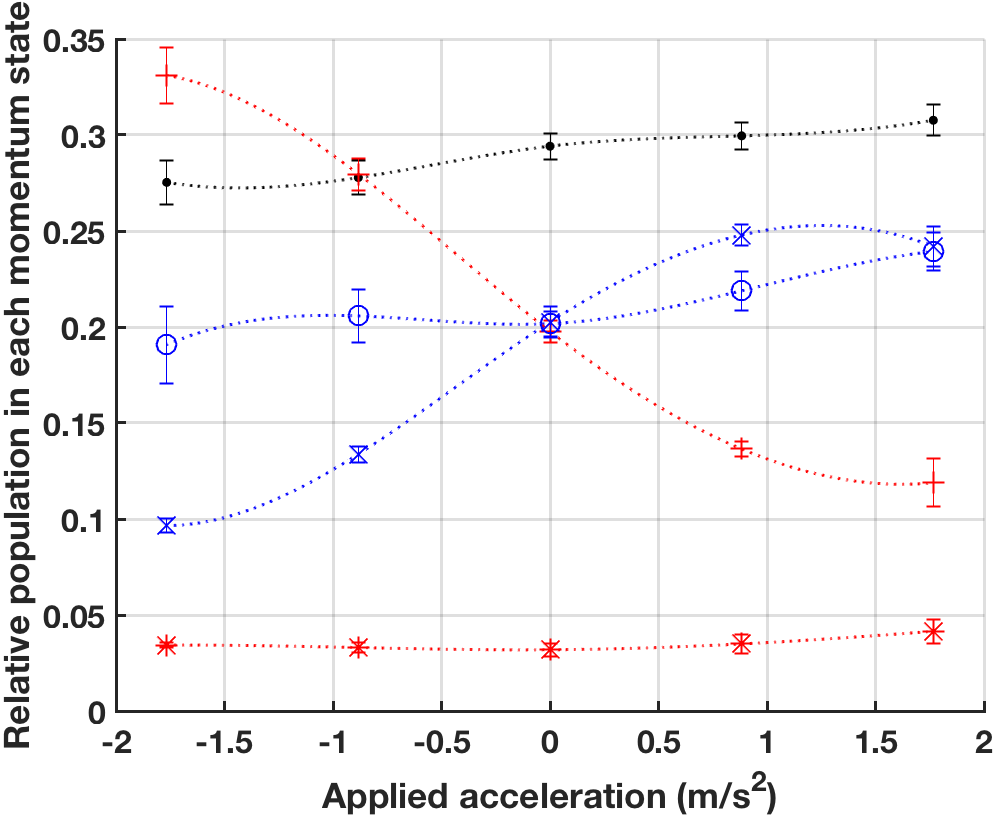}
\caption{\label{fig:direction} Here and in Fig. \ref{fig:sprec_bias}, the momentum population of the atoms after the $T_\mathrm{I} = 2$~ms  interferometer sequence as a function of the applied acceleration signal. Atoms in the $2n\hbar k_\mathrm{L}$ state are denoted by open blue circles ($n = -2$), blue crosses ($n = -1$), black dots ($n = 0$), red plusses ($n = 1$) and red asterisks ($n = 2$). As the applied signal is varied away from zero, we can distinguish positive and a negative signals. The dotted lines are cubic spline fits to guide the eye.}
\end{figure}

We can calibrate the interferometer response to a signal by recording how the final state of the interferometer after shaking changes with the applied signal. Because the lattice shaking breaks the symmetry of the system \cite{Anderson2017}, we can determine the sign of an applied signal. We measure the variation of the atoms' final momentum state after the interferometry sequence, as shown in Fig. \ref{fig:direction}. The data show that the final state after an acceleration $a$ is applied is distinct from the final state after an acceleration $-a$ is applied. This ability to distinguish the signal direction differentiates our interferometer from the typical light-pulse atom interferometer where the atom population varies cosinusoidally between two states.

\begin{figure}[h!]
\includegraphics[scale = .72]{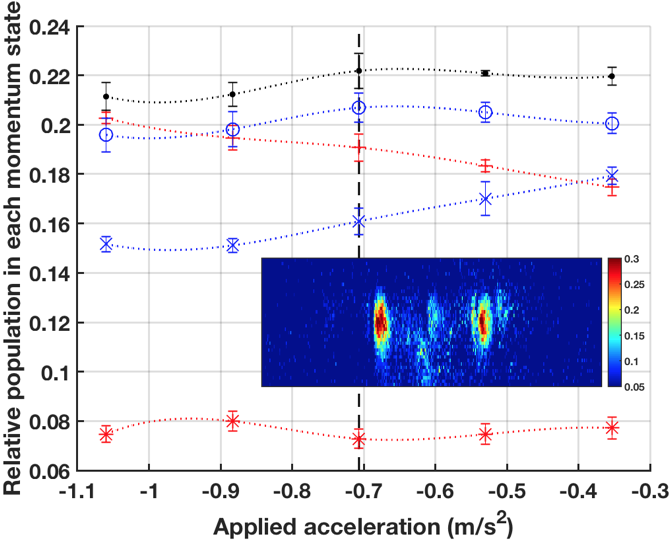}
\caption{\label{fig:sprec_bias} Plot of the momentum state variation as a function of applied acceleration with the biased interferometer, showing variation of the final state as the acceleration is varied around the optimized bias value of $a_\mathrm{bias} = -0.71$ m/s$^2$ (black dashed line). Data points and splines colored as in Fig. \ref{fig:direction}. (inset) An experimental image of the optimized split state in the biased interferometer. OD is indicated by the colorbar on the right.}
\end{figure}

Finally we show steps towards the tunability of the interferometer transfer function \cite{Anderson2017}. We optimize the interferometer in the standard Michelson configuration but add a bias signal $a_\mathrm{bias} = -0.71$~m/s$^2$ during optimization. We then measure the atoms' final  recombined momentum state after the addition of signals $a_\mathrm{bias} \pm \Delta a$, as shown in Fig. \ref{fig:sprec_bias}. From this data, we see that we can distinguish the sign of $\Delta a$ by observing the final state of the atoms. Further extensions of this work include increasing the magnitude of $a_\mathrm{bias}$ and optimization of the interferometer to a AC-varying signal, as predicted in Ref. \cite{Anderson2017}. This will allow the interferometer to be optimized for sensitivity to any signal of interest.

In conclusion we have presented experimental results of interferometry using atoms trapped in an optical lattice, showing that shaken lattice interferometry scales as $T_\mathrm{I}^2$. The sign of the applied signal may be measured, and the interferometer may be optimized in the presence of a bias signal. We show that the limitation on our interferometer sensitivity is set by imaging noise, which may be mitigated with some straightforward experimental improvements. Atom stability in the lattice can be improved by the use of common intensity stabilization techniques allowing for longer interrogation times limited only by photon scattering rates and collisions with background particles. Thus by improving imaging and the stability of the lattice laser, the interferometer interrogation time and sensitivity limit can be improved. Finally, it is straightforward to expand this system to work in a three-dimensional lattice system, paving the way towards a sensor capable of simultaneously measuring accelerations along three axes.

\begin{acknowledgments}
The authors would like to acknowledge funding from the NSF PFC under Grant No. 1125844 and Northrop Grumman Corporation.
\end{acknowledgments}

\bibliography{20180127_SLI_expt}

\end{document}